**Title**
Large-area, all-solid and flexible electric double layer capacitors based on CNT fiber electrodes and polymer electrolytes.

*Author(s), and Corresponding Author(s)* Evgeny Senokos, Víctor Reguero, Laura Cabana, Jesus Palma, Rebeca Marcilla*, Juan Jose Vilatela*

E. Senokos, Dr. J. Palma, Dr. R. Marcilla
IMDEA Energy Institute
Avda. Ramón de la Sagra 3, Móstoles, 28935 Madrid, Spain
E-mail: rebeca.marcilla@imdea.org
E. Senokos, V. Reguero, L. Cabana, Dr. J. J. Vilatela
IMDEA Materials Institute
C/ Eric Kandel, 2, Getafe, 28906 Madrid, Spain.
E-mail: juanjose.vilatela@imdea.org

This work presents a scalable method to produce robust all-solid electric double layer capacitors (EDLCs), compatible with roll-to-roll processes and structural laminate composite fabrication. It consists in sandwiching and pressing an ionic liquid (IL) based polymer electrolyte membrane between two CNT fiber sheet electrodes at room temperature, and laminating with ordinary plastic film. This fabrication method is demonstrated by assembling large area devices of up to 100 cm$^2$ with electrodes fabricated in-house, as well as with commercial CNT fiber sheets. Free-standing flexible devices operating at 3.5 V exhibited 28 F g$^{-1}$ of specific capacitance, 11.4 Wh kg$^{-1}$ of energy density and 46 kW kg$^{-1}$ of power density. These values are nearly identical to control samples with pure ionic liquid. The solid EDLC could be repeatedly bent and folded 180° without degradation of their properties, with a reversible 25% increase in energy density in the bent state. Devices produced using CNT fiber electrodes with a higher degree of orientation and therefore better mechanical properties showed similar electrochemical properties combined with composite specific strength and modulus of 39 MPa/SG and 577 MPa/SG for a fiber mass fraction of 11 wt.%, similar to a structural thermoplastic and with higher specific strength than copper.

**1. Introduction**

There is an ever-increasing interest in energy storage devices that not only store energy but which are also flexible, stretchable, or structural.[1–7] These mechanical requirements demand new electrode materials and architectures. Most conventional supercapacitors, for example, consists of a layer of activated carbon paste deposited onto a metallic current collector, a liquid electrolyte and a metallic casing to contain it. The use of metallic sheets in such configuration not only adds substantial weight to the final device but hinders the mechanical functions described above. Similarly, most activated carbon, produced by carbonization of waste natural fiber, is a brittle powder incompatible with large strains even when embedded in polymer matrices.[8–10]

In this context, nanocarbons have been a popular building block for electrodes capable of withstanding large deformation and often stresses. Carbon nanotubes and graphene possess high surface area, high electrical conductivity (when metallic and highly crystalline) and unrivalled mechanical properties. Equally important, they can be assembled into large-area percolating networks with high porosity, typically as thin membranes. The membranes are highly flexible, a consequence of the small thickness of the network building blocks (bending stress scales with thickness to the power of 4). Thus, for example, reports on flexible electric double layer capacitors (EDLCs) based on carbon nanotubes indicate capacitance in the range of 7 - 203 F $g^{-1}$.[5,11–15]

In spite of this success, there are still clear challenges in the development of these capacitors in terms of: fabrication of large-area devices using simple integration routes, use of industrially-produced CNTs, obtaining mechanical properties beyond flexibility, and in general in comparing properties of different devices and assessing the extent to which the properties of the individual building blocks are efficiently exploited. In the case of fabrication, for example, the vast majority of these devices are produced by solvent-based processes. Typically, this involves producing a dispersion of nanocarbons in a solvent or a polymer solution and then depositing it as a thin film by drop casting[16], blade coating[17], spin coating[18] and spray coating[19] or similar methods. Dispersing CNT invariably leads to the use of chemical functionalization routes that thus degrade their low-dimensional properties and usually restrict the choice of CNTs to multiwall. Although the resulting materials are flexible, their mechanical properties are still very weak.

In this work we present a simple method to fabricate all-solid EDLCs, by simply pressing of sheets of CNT fibers and a polymeric electrolyte membrane. The process is demonstrated for devices of different shapes and size up to 100 $cm^2$. We then compare their electrochemical properties to those obtained from two- and three-electrode cells using pure ionic liquid in order to evaluate the efficiency of the assembly process. We also fabricate devices using commercial CNT fiber sheets from three different suppliers and compare them with material produced in-house. The comparison shows that for nanocarbon-based EDLCs power density scales as *1/mass loading* as a consequence of charging time increasing linearly with electrode thickness. We then present evidence of electrochemical properties of the solid devices being retained, and even improved, after successive bending and folding. Tensile properties of the CNT sheet/polymer electrolyte composites are also presented.

## 3. Results and discussions

**Basic structure and electrochemical properties of CNT fiber electrodes**

The electrodes used in this study consist of sheets of CNT fibers produced by drawing an aerogel of CNT directly from the gas phase during their growth by chemical vapour deposition (CVD).[20] This method enables production of large-area uniform samples in semi-industrial quantities,[21] while also offering the possibility to tailor their structure[22] and molecular composition[23,24] and thus preserve some of the low-dimensional properties of the constituent CNTs on a macroscopic scale.[25] Photographs of a typical electrodes produced in our laboratory are presented in **Figure 1a**-b (details in Experimental section). The electrode consists of a network of long CNTs associated in bundles but imperfectly packed and thus giving rise to a large porosity (Figure 1c). The first part of this study deals with sheets of CNT with a low degree of orientation, produced under conditions for high-throughput[22] that thus resemble those used in commercial facilities. The CNTs are highly graphitised and have few layers (<5) (Figure 1d), leading to a specific surface area of 240 $m^2$ $g^{-1}$, and a predominance of meso and macropores (Figure 1S, Supporting Information).

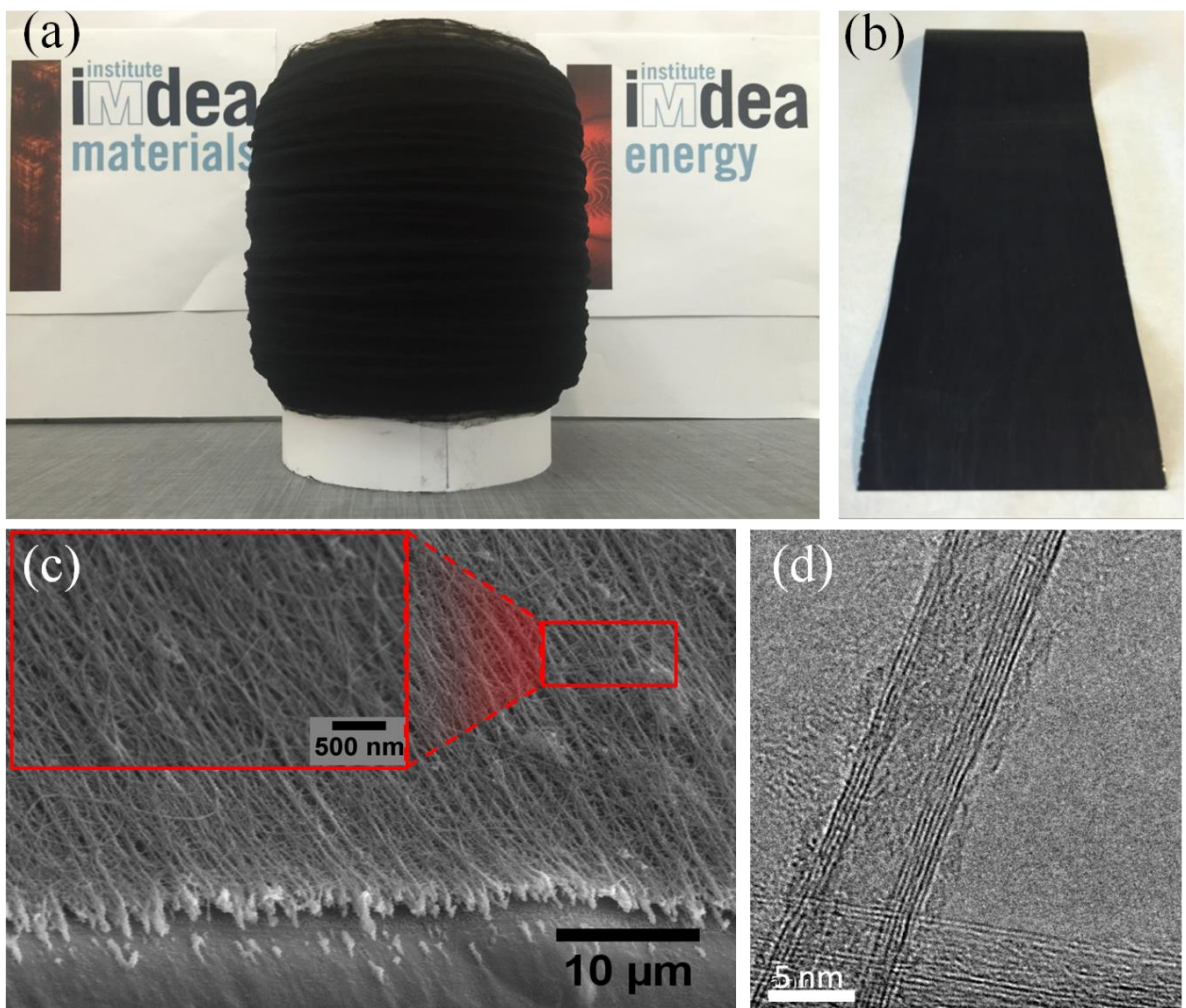

**Figure 1.** CNT fiber sheets used as electrodes. a) 1km of CNT fibers continuously spun and deposited on aluminium foil. b) CNT fibers sample after densification with acetone, c) SEM image showing the porous structure of CNT fibers. d) TEM image showing that the CNTs have few layers and a high degree of graphitization.

The mass loading (areal density) of a CNT fiber sheet electrode produced by this method can be easily adjusted by simply varying the time that fibers are deposited onto the winder. Because there are no major changes in volumetric density as the CNT fiber is continuous laid down on the winder, electrode thickness can be taken as proportional to mass loading. This opens the possibility to first study and then optimize parameters of the electrode assembly that could ultimately affect the performance of a final device.

Figure 2 presents electrochemical results from galvanostatic charge-discharge (CD) measurements at 2 mA cm$^{-2}$ on symmetric CNT fiber electrodes in pure PYR$_{14}$TFSI ionic liquid (IL), using a commercial polypropylene separator and for different electrode mass loadings (i.e. thickness). The Ragone plot (Figure 2a) shows that whereas energy density remains fairly constant at around 11 Wh kg$^{-1}$ at lower current densities, power density drops rapidly as electrode thickness is increased. Specific capacitance (C$_s$), calculated from the slope of the discharge profile and normalized by mass loading, is also independent of electrode thickness, at around 23 F g$^{-1}$ (Figure 2b). Similarly, equivalent series resistance (ESR) extracted from the ohmic drop is approximately constant (35 Ω cm$^2$) (Figure 2c). In contrast, charge-discharge times depend strongly on electrode thickness. Figure 2d shows that discharge time is proportional to mass loading and thus, specific power is inversely proportional to mass loading.

This is in fact the predicted behavior for a highly conductive electrode when all surface area is ionically accessible and micropore transport effects are neglected. For an ideal symmetric EDLC under a constant applied current the discharge time can be expressed as[26]

$$t_D = \frac{V - V_0 - IR_\infty}{2I} \times C_s \rho_V L$$

Where $V$ is the voltage across the cell, $V_0$ is the initial potential, $I$ is the current, $R_\infty$ is the electrode resistance, $C_s$ is the gravimetric capacitance, $\rho_V$ is the volumetric density and $L$ is the electrode thickness. The results in Figure 2 indicate that electrode thickness limits power per electrode mass through an inverse relationship and that thicker electrodes are inherently slower. This might be a simple dependence, but nevertheless it is evident that it has important implications for any real device of practical relevance, as well as to compare values of power

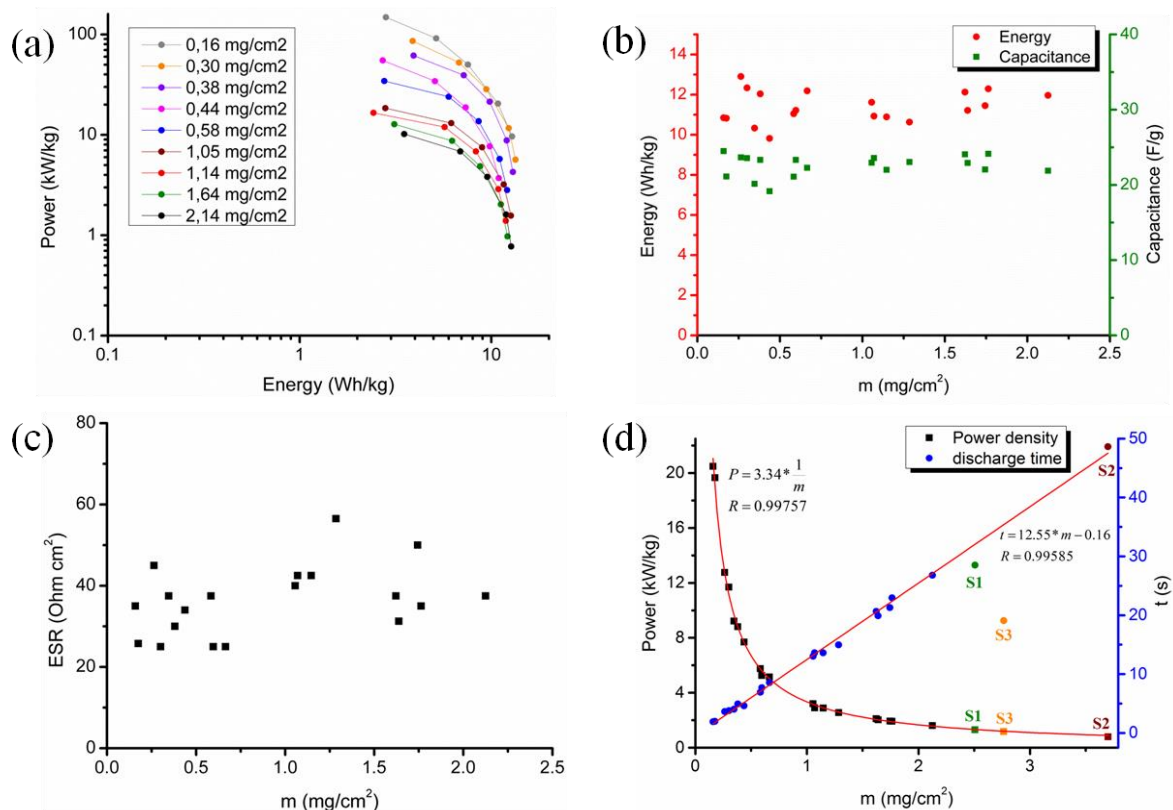

**Figure 2.** Electrochemical properties of CNT fiber sheet electrodes as a function of electrode mass loading (i.e. thickness) obtained from charge-discharge measured at 2 mA cm$^{-2}$. a) Ragone plots of IL based EDLCs with different thickness of CNT fiber electrodes, b) Energy density, specific capacitance and c) ESR does not change significantly for different electrode thickness. d) Discharge time increases linearly with electrode mass loading and thus power density depends inversely on electrode thickness.

density reported in the literature, which can appear high only on account of sample dimensions.

Figure 2d also includes results from charge-discharge (CD) of coin-cell devices with commercial CNT fiber sheets. Commercial CNT fiber samples from suppliers 1 (S1, mass loading 2,51 mg cm$^{-2}$) and 2 (S2, mass loading 3.70 mg cm$^{-2}$) fall approximately on the same line as those produced in-house, while this particular sample from the supplier 3 (S3, mass loading 2.76 mg cm$^{-2}$) has a shorter CD time and it follows the same figure of merit of *power density times mass loading*. This is evidenced by the CD profile in Figure 2Sa (Supporting Information) where the lower capacitance and energy density of the commercial carbon material obtained from supplier 3 were observed. These results are in line with cyclic voltammetry (CV) measurements of CNT fiber in IL and 3-electrode electrochemical cells, included in Supporting Information (Figure 2Sb, Supporting Information). At 5 mV s$^{-1}$ and 3.5 V, for example, materials produced in-house, S1 and S2 have a specific capacitance around 31 F g$^{-1}$ whereas that for the S3 sample is only 24 F g$^{-1}$. The observed difference in electrochemical performance is explained by compositional differences between CNT fibres from the various sources. The sample from supplier 3 has a significantly higher fraction of impurities, for example 41 wt.% residual catalyst, which reduces the gravimetric surface area and therefore capacitance of the sample (Figure 3S, Supporting Information). CV at high voltages also showed that all samples exhibited a nonlinear increase of differential capacitance leading to voltammograms with a butterfly shape. This is due to a significant quantum (chemical) capacitance in the samples as a consequence of their constituent CNTs being highly graphitized and having few layers [25]. Chemical capacitance arises in materials with a low density of states DOS near the Fermi level. In addition to an electrostatic geometric capacitance, they present a second capacitive contribution ("in series"), corresponding to changes in electrochemical potential as a result of higher occupation of energy levels.[27] In the case of nanocarbons, this occurs when they their low-dimensional density of states resembles that of a zero-gap semiconductor, as opposed to that of graphite. This requires a high degree of graphitization and few layers in the constituent building block.[28,29] For bulk samples of CNTs containing a distribution of diameters and chiral angles, the joint DOS resulting from superposition of individual DOS retains a low density of states near the Fermi level, although it loses single-molecule features such as van Hoff singularities and leads to a smooth almost linear lineshape.[30]

The properties of CNT fiber sheet electrodes presented above correspond to tests in coin-cell configuration using pure ionic liquid electrolyte and to some extent reflect intrinsic properties of the electrode materials. They are a good basis to later evaluate the performance of all-solid supercapacitors and identify the effects of replacing the IL with a solid polymer electrolyte that also takes the role of separator.

**Assembling and electrochemical characterization of all-solid EDLC based on CNT fibers and polymer electrolyte membrane.**

Conventional methods of EDLC fabrication are based on sandwiching electrodes impregnated with electrolyte and a separator at high pressure inside a metallic casing or in a pouch cell. The processes normally involves over 15 stages, including forming a slurry of carbon, coating it on a current collector, assembling into a cell, electrolyte injection, vacuum standing, sealing, etc. To expedite the process to fabricate free-standing EDLC devices while also making it compatible with roll-to-roll techniques, we have developed a simple process based on room-temperature pressing of two CNT fiber sheets and a polymer electrolyte membrane, and a further step of lamination in plastic using an ordinary laminating machine.

A scheme of the assembly of all-solid EDLC is presented in Figure 3a. The first step consists in the deposition of CNT fibers onto a metallic sheet, whose main purpose is to provide a stiff support for the electrodes and can be conveniently used as current collector. However, it is worth mentioning that this component is ultimately redundant on account of the high longitudinal conductivity of the CNT fibers $> 3.5 \times 10^3$ S cm$^{-1}$, which could thus also take the role of current collectors (Figure 4S). The EDLC structure is assembled by sandwiching a pre-formed membrane of polymer electrolyte between two CNT fiber sheet electrodes. The membrane is produced by doctor blading a viscous solution of PVDF-co-HFP and an IL in a common solvent. For this work we chose PYR$_{14}$TFSI as IL electrolyte due to its wide electrochemical window and high ionic conductivity[27,28], and PVDF-co-HFP as polymeric matrix because of its excellent mechanical and thermal stability and semi-crystalline nature[29,30]. The mixture of PYR$_{14}$TFSI ionic liquid, PVDF-HFP copolymer and Li salt has been investigated as electrolyte in conventional lithium-ion batteries and, in combination with CNT fiber cathode in lithium-oxygen battery [31]. Although PVDF-HFP is a common polymer component of PE used with several ionic liquids for solid EDLC[36,37], its combination with PYR$_{14}$TFSI has not been investigated so far. Uniform and transparent membranes with no microphase separation were produced with a concentration of IL in PVDF-HFP between 50 and 70% (Figure 3b). At 60 wt.% the membranes were found to have the optimum balance of mechanical robustness and ionic conductivity and this composition was thus used to assemble all the solid devices in this study. Their room-temperature conductivity is $4 \times 10^{-4}$ S cm$^{-1}$, compared with $2.7 \times 10^{-3}$ S cm$^{-1}$ for the pure IL.

The different electrode and electrolyte layers of the EDLC are consolidated at room temperature simply by applying a small pressure. This method provides sufficiently good impregnation of the soft polymer electrolyte (PE) membrane into the porous CNT fiber structure. Figure 3c presents an electron micrograph of the cross section of a CNT fiber sheet/PE membrane/CNT fiber sheet composite structure extracted from a fully operational device. The interfacial regions show extensive infiltration of the polymeric phase into the fiber structure and no evidence of large gaps or delaminated areas. The membrane used is substantially thicker (38 µm) than the CNT fiber sheets (7 µm). This ensured reproducibility in the fabrication of large area devices, but it is clearly only a starting point for further optimization. Adequate impregnation of the membrane into the nanostructured electrodes was confirmed by comparison of CD measurements on EDLC devices assembled with pure liquid and solid electrolytes in a 2-electrode Swagelok cell (Figure 5S, Supporting Information). These results showed that capacitance and energy densities were only slightly lower to those obtained with pure IL confirming the adequate filling of the voids of CNT fiber with the PE. The final fabrication step consists in encapsulating the EDLC using a conventional plastic laminate machine. The laminated configuration protects the device from humidity absorption, keeps the structure under a small compressive stress normal to the EDLC layers, but preserves the high flexibility of the materials in the device (discussed below). Figure 4a,b presents results from CD for all-solid EDLCs of 4.5 cm$^2$ and a mass loading of 1.15 mg cm$^{-2}$, corresponding to a thickness of a single electrode around 14 μm. It compares results before lamination and 90 hours after lamination. The small compressive stress applied during lamination has a positive effect on capacitance and energy density of the devices, as previously observed in similar systems discussed in the literature.[34,35] The data show that capacitance reaches 28 F g$^{-1}$ at 1 mA cm$^{-2}$, 86% of the value obtained in IL at the same current density, confirming the efficient exploitation of electrode properties using the simple fabrication method discussed above. This leads to power and energy densities as high as 46 kW kg$^{-1}$ and 11.4 Wh kg$^{-1}$, respectively, together with a coulombic efficiency > 97%. Long-term stability of assembled devices (both before and after lamination) was investigated by conducting 10000 charge-discharge cycles from 0V to 3.5V at 5 mA cm$^{-2}$. Figure 4c shows that capacitance retention after cycling was higher than 96% demonstrating the excellent stability of the EDLC over cycling.

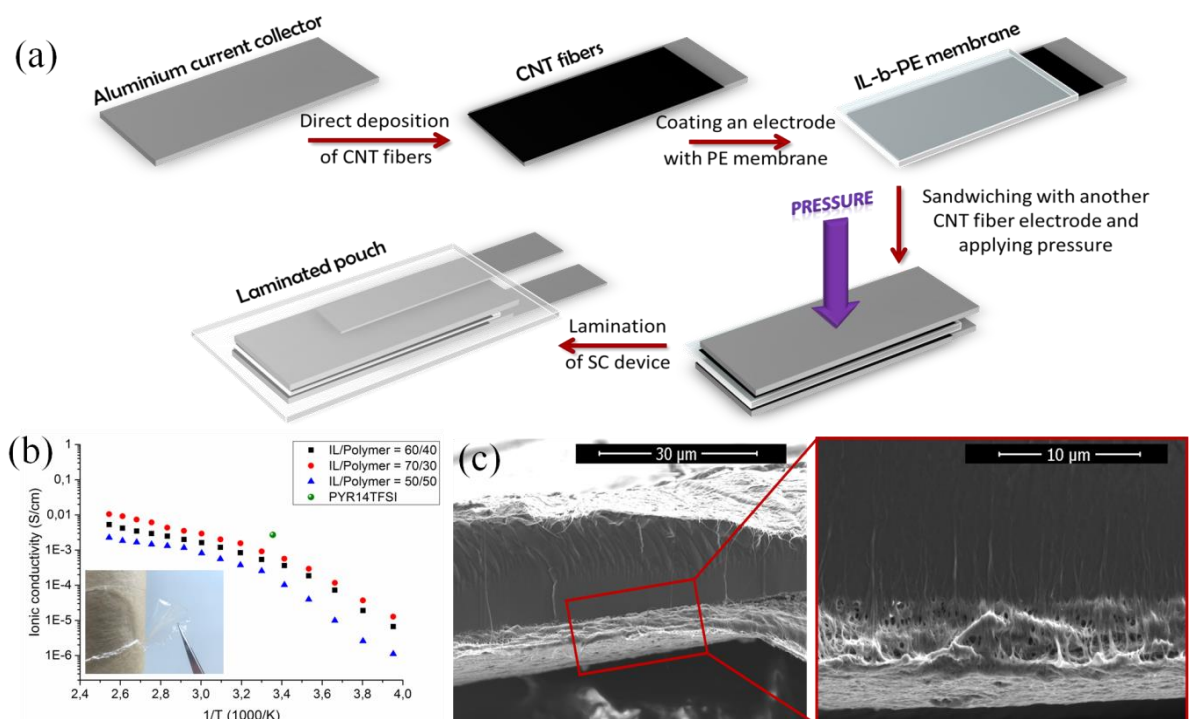

**Figure 3.** Assembly of all-solid EDLC. a) Scheme of the fabrication of devices by pressing a polymer electrolyte membrane between two CNT fiber sheet electrodes and laminating. b) Arrhenius plot of ionic conductivity measured for PE with different IL/PVDF-HFP mass ratio compared to pure IL ionic conductivity at 25ºC and photo of transparent freestanding PE membrane (inset). c) Cross-section SEM image of CNT/PE/CNT composite in a solid EDLC.

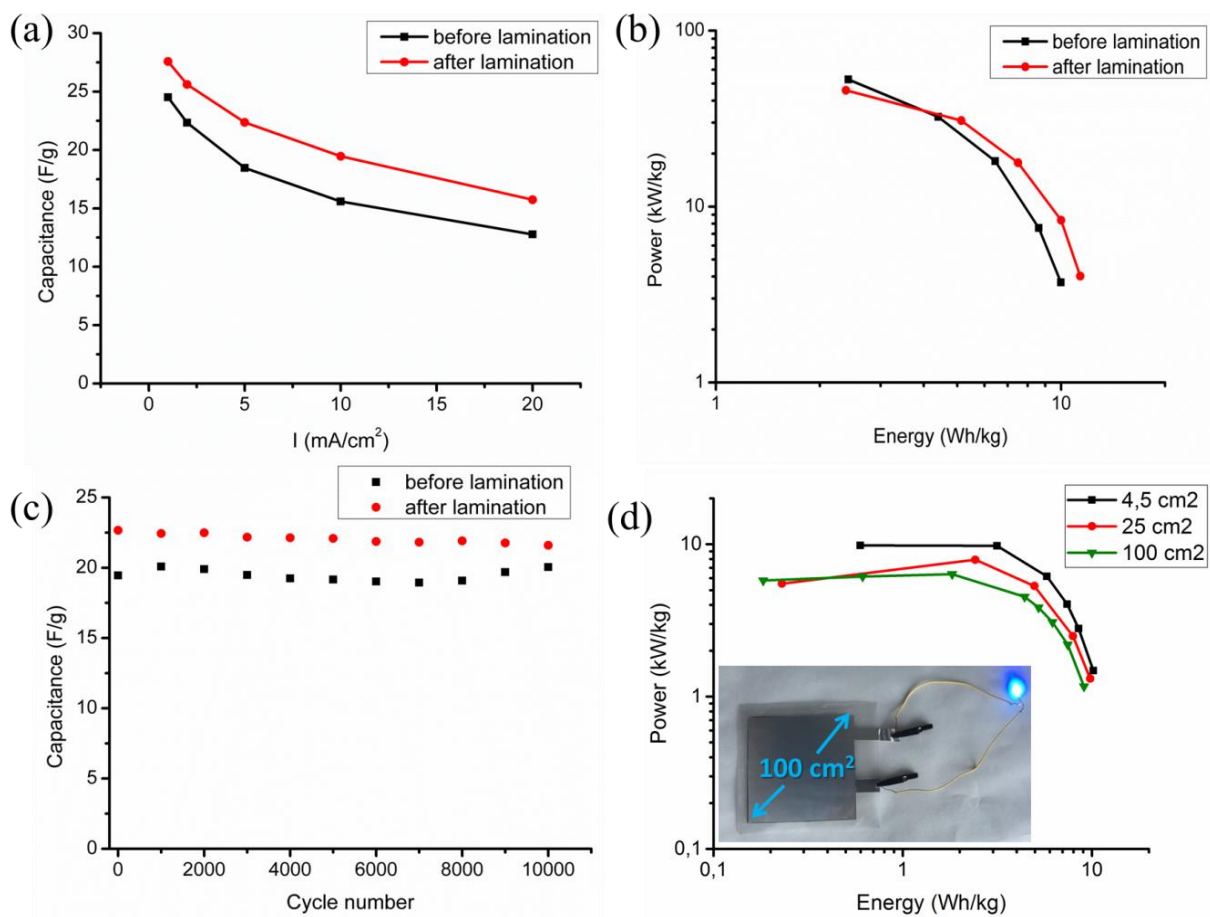

**Figure 4.** Electrochemical properties of devices. a) Capacitance, b) Ragone plot and c) Capacitance retention during 10000 cycles at 5 mA cm$^{-2}$ and 3.5V of applied voltage obtained for all-solid EDLC before lamination and after lamination in free-standing state, d) Ragone plot of 4.5, 25 and 100 cm$^2$ laminated devices and photo of freestanding solid EDLC of 100 cm$^2$ lighting a blue LED (inset).

The EDLC assembly process could be easily scaled up as well as using commercial CNT fiber sheets. Figure 6S (Supporting information) shows photographs of all-solid 4.5 cm$^2$ EDLC devices assembled with CNT fibers purchased to different suppliers lighting red LEDs during their discharge. Figure 4d shows a Ragone plot for in-house EDLC of different size and shape confirming that the scaling up process does not have a significant impact on the electrochemical properties (energy density is similar for all the cases and differences in power density were attributed to the different electrode mass loading from one EDLC to another). Rectangular 100 cm$^2$ all-solid EDLC (inset Figure 4d) were successfully produced following the assembly stages described above. These scaled-up devices (100 cm$^2$) showed good reproducibility (Figure 7S, Supporting Information) and similar electrochemical properties to smaller devices. For 100 cm$^2$ devices, capacitance was around 22 F g$^{-1}$, with energy density reaching 8.8 Wh kg$^{-1}$ and power density 4 kW kg$^{-1}$ for a mass loading of 2.4 mg cm$^{-2}$. Power density values for solid state devices confirmed the same reciprocal dependence on electrode mass loading observed before in IL (Figure 8S, Supporting Information), which itself implies no degradation of properties when increasing sample size.

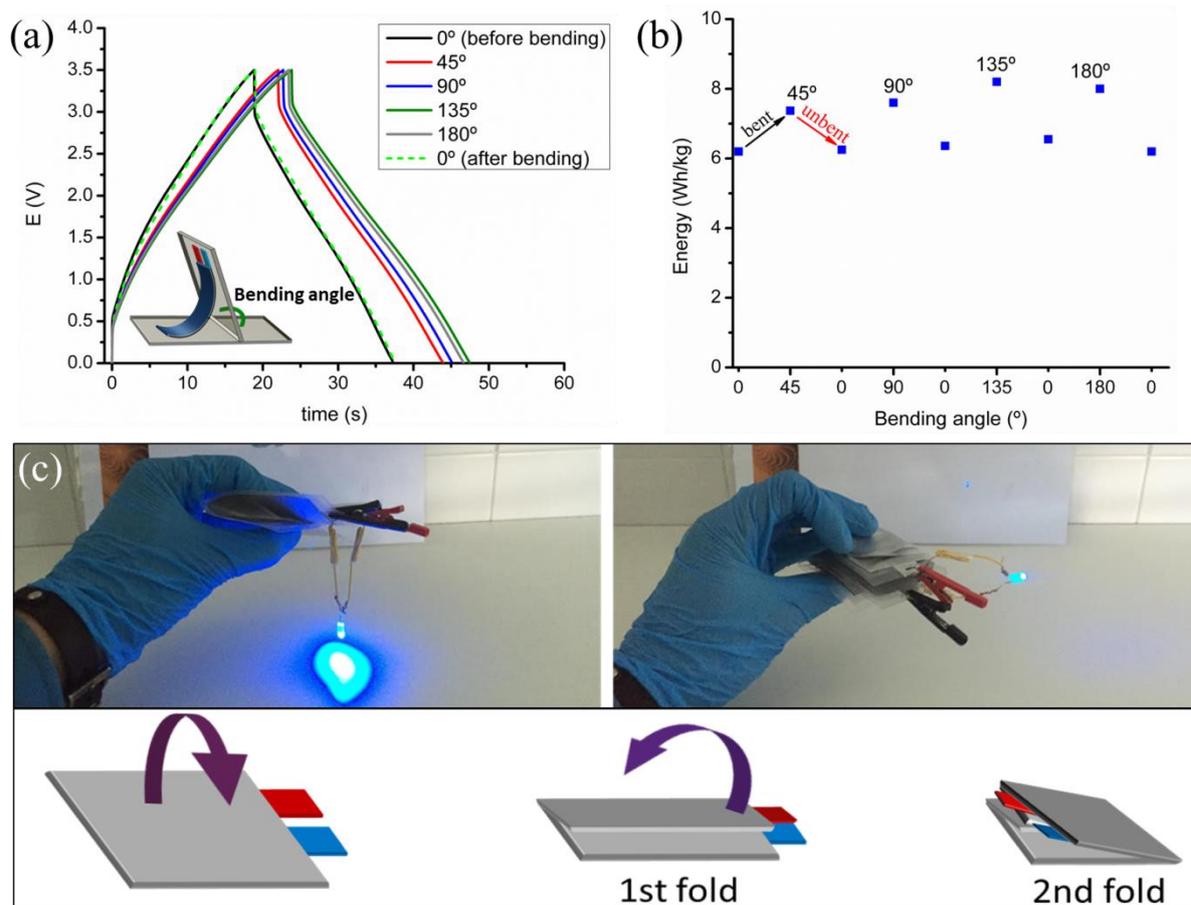

**Figure 5.** Operation of devices under flexural deformation. a) Charge-discharge profiles of EDLC and b) calculated energy density obtained at 2 mA cm$^{-2}$ for different bending angles from 0º to 180º. c) Photographs and schematic of a 100 cm$^2$ laminated EDLC device powering a blue LED after multiple folding.

All EDLC devices produced in this study show a high flexibility, without deterioration of electrochemical properties after several bending cycles. Galvanostatic charge discharge measurements were performed on 100cm$^2$ samples at different deflection angles from 45 to 180º. To ensure that electrochemical measurements did not correspond to a transient state associated with the viscoelasticity of the PE, all measurements were carried out 1 hour after bending/releasing. The results are presented in Figure 5a,b. Galvanostatic CD curves show that bending the devices slightly decreases the ohmic drop (decreasing ESR) while keeping appropriate coulombic efficiency. Very importantly, after unloading the device to the unstrained state, even after being fully folded, it recovered its initial properties. Power density was approximately constant throughout the bending cycles, whereas energy density increased substantially in the bent state, but then went back to its original value after unloading (Figure 5b). This is most likely a consequence of the bending stress locally improving polymer infiltration which would have the effect of increasing energy density but not power density (Figure 9S, Supporting Information), but we do not entirely rule out the possibility of piezoelectric effects also making a contribution.[40] As a further demonstration of device flexibility we show a device powering a blue LED after folding it on itself twice (Figure 5c).

A unique feature of CNT fiber-based electrodes is their combination of high surface area and high mechanical properties,[37] which could be exploited to produce devices which are not only flexible but also have structural properties. The long term goal is to produce structural composites with energy storing capabilities that thus combine two functions and can lead to a net weight reduction, for example in components used in transport.[42–44] As a step in this direction, we have produced electrodes based on CNT fibers purposely synthesized to have a high degree of CNT orientation parallel to each other and to the fiber axis, and which thus lead to higher strength and stiffness than the high-throughput grade (see experimental details). Their electrochemical properties are very similar to high-throughput grade CNT fibers, and lead to all-solid devices with energy density in the range of 9.7 - 10.1 Wh kg$^{-1}$ and a power density of 8.7 – 18.3 kW kg$^{-1}$ for a mass loading of around 1.2-1.7 mg cm$^{-2}$ (Figure 6a). A few of these capacitors were subsequently subjected to mechanical testing. The CNT fiber sheet/PE membrane/CNT fiber sheet EDLC structure was removed from the plastic laminate and current collectors and tested in tension. Figure 6b shows a photograph of a typical device as well as examples of the stress-strain curves. The maximum tensile strength and modulus measured were 53 MPa and 790 MPa, compared with 2.8MPa and 5MPa for the polymer electrolyte membrane. In specific units these composite values correspond to 39 MPa/SG and 577 MPa/SG, while specific toughness (energy-to-break) calculated from the area of stress-strain curve reached 3.7 J g$^{-1}$. The value of specific strength is for example, higher than that of a 102 copper alloy (22-36 MPa/SG). But ultimately, these results should be taken as a starting point for further improvements on mechanical properties. We note that the fiber volume fraction in these composites is below 14 vol.%, whereas traditional fiber-reinforced polymer composites have typically volume fractions above 50%. Furthermore, we note that these electrodes have not been optimized for cooperative load bearing between individual CNT fiber filaments.

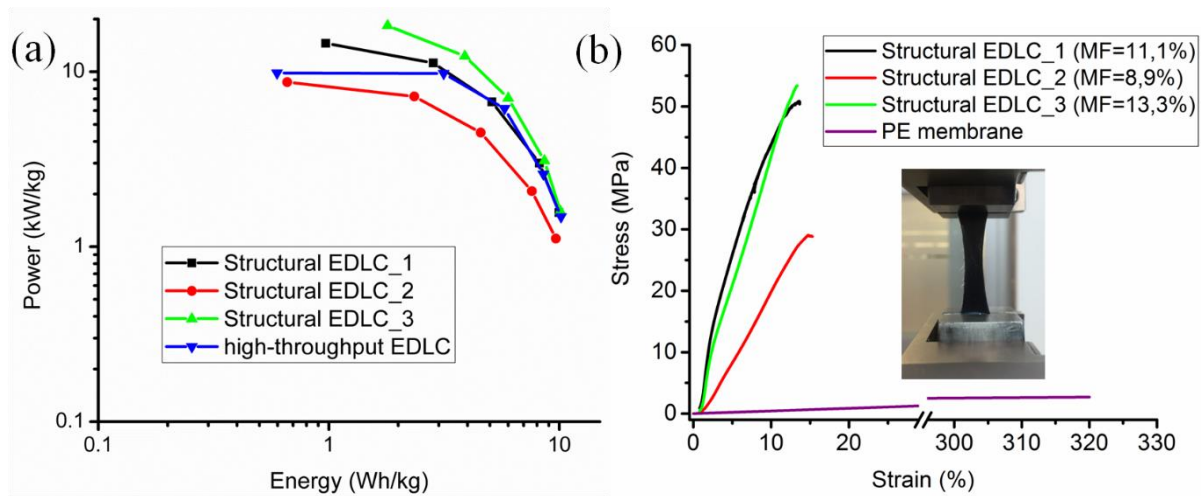

**Figure 6.** a) Ragone plot comparing electrochemical properties of free-standing EDLCs based on highly-oriented (structural EDLC_1, 2 and 3) and high-throughput (high-throughput EDLC) CNT fibers produced in house b) Stress-strain curves obtained for CNT fibers/PE composites with different mass fraction (MF) of the carbon material previously used as EDLCs.

Finally, Table 1 presents a summary of electrochemical and mechanical properties of reported flexible all-solid EDLC, for reference. The comparison highlights the fact that most literature reports are on very small samples compared with real devices, often with active material mass loadings of tens of micrograms/cm$^2$. The exceptions are this work and reports by Imperial and co-workers[53,56], both corresponding to multifunctional materials that combine energy storage and structural properties.

**Table 1.** Comparison of flexible all-solid EDLCs

| Electrode material | Electrolyte | Capacitance of the device, F g$^{-1}$ [mF cm$^{-2}$] | Energy, Wh kg$^{-1}$ [µWh cm$^{-2}$] | Power, kW kg$^{-1}$ [mW cm$^{-2}$] | a)Strength of SC composite, MPa g$^{-1}$cm$^3$ | a)Modulus of SC composite, MPa g$^{-1}$cm$^3$ | Size of device, cm$^2$ |
|---|---|---|---|---|---|---|---|
| CNT fibers (this work) | [PYR$_{14}$][TFSI]-(PVDF-HFP) | 28 | 11.4 | 46 | 53 (tensile) | 790 (tensile) | 100 |
| Activated carbon/CNTs[47] | [BMIM][BF$_4$] - (ETPTA) | N/A [100] | 10.3 | 3 | N/A | N/A | micro SC |
| CNTs[48] | [EMIM][TFSI]-(PS–PEO–PS) | 57 [0.34] | 17.5$^{b)}$ [0.42] | 1.7 $^{b)}$ [35] | N/A | N/A | 1 |
| Reduced graphene oxide (RGO)/ Polyaniline (PANI)/MWCNT/IL[49] | [EMIM][BF$_4$] | 154 | 12 | 17.2 | N/A | N/A | 0.3 |
| Graphene hydrogel[5] | H$_2$SO$_4$-PVA | 186 [372] | 6.5 | 5 | N/A | N/A | 1.75 $^{c)}$ |
| RGO-poly(3,4-ethylenedioxythiophene) polystyrene sulfonate (PEDOT/PSS)[50] | H$_3$PO$_4$-PVA | 52.7 | 2.83 | 3.59 | N/A | N/A | 1.6 (30)$^{d)}$ |
| Celulose nanofiber/RGO aerogel[15] | H$_2$SO$_4$-PVA | 207 | 6 | 4.65 [15.5]$^{e)}$ | N/A | N/A | 2 |
| CNTs/Bacterial Nanocellulose[52] | [EMIM][TFSI]-(PS–PEO–PS) | 50.5 | 15.5 | 14$^{f)}$ | N/A | N/A | 1 |
| CNT/paper[53] | [EMIM][TFSI]-silica | 134.6 | 41$^{g)}$ | 40$^{g)}$ | N/A | N/A | 1 |
| Carbon aerogel modified Carbon Fiber (CF) Fabrics[45] | [EMIM][TFSI]-PEGDGE | 0.6 | 8.4×10$^{-4}$ | 3.3×10$^{-5}$ | 5 (shear) | 519 (shear) | 59.5 |
| CNT/Carbon cloth[53] | H$_3$PO$_4$-PVA | 106.1 [38.75] | 6.6 [2.4]$^{h)}$ | 52 [19]$^{h)}$ | N/A | N/A | 1 |
| 3D N-doped activated CNF/bacterial cellulose[54] | H$_2$SO$_4$-PVA | 175 | 6.1 | 50 | N/A | N/A | 2.4 |
| Activated CF[46] | LiTFSI + [EMIM][TFSI]-PEGDGE | 0.052 | 1.4×10$^{-3}$ | 2.7×10$^{-3}$ | 4.58 (compressive) | 11000 (tensile) | 25 (323) |
| P−Si layers[55] | [EMIM][BF$_4$]-PEO | N/A | 9 | 8 | N/A | N/A | N/A (<5) |
| Polyacrylonitrile (PAN) - based CF[57] | LiTFSI-(25% SR494+75% CD552)$^{i)}$ | 0.093 | 0.021 | 1.5×10$^{-4}$ | 485 (tensile) | 12000 (tensile), 310 (shear) | 25 |
| RGO films[58] | $H_2SO_4$-Polyimide | 53.2 [71][j] | 7.35 [9.8][j] | 30 [40][j] | N/A | N/A | 1.5 |
| Highly aligned CNTs[59] | $H_3PO_4$-PVA | 29.2 | 2.4 | 0.9 | N/A | N/A | 1.2 - 3 |

[a] Mechanical properties are measured without current collector or laminating pouch. [b] Recalculated from areal energy and power assuming 12 mg cm$^{-2}$ of mass loading for 1 electrode. [c] Calculated from the dried weight and mass density of electrode material, [d] Large area device was not characterized. [e] Recalculated from areal power density assuming 3.33 mg cm$^{-2}$ of mass loading for electrode material. [f] Power density value estimated from Ragone plot, [g] Energy and power values obtained from Ragone plot. [h] Energy and power density values recalculated assuming 0.36 mg cm$^{-2}$. [i] The polymer consists of 25% SR494 (Sartomer) ethoxylated pentaerythritol tetraacrylate and 75% CD552, a methoxy polyethylene glycol 550 monomethacrylate. [j] Values of specific gravimetric capacitance, energy and power are recalculated assuming 1 mg cm$^{-2}$ of one electrode.

## 3. Conclusions

This work has introduced a simple method to produce all-solid EDLCs based on electrodes made up of CNT fiber sheets. The EDLCs are assembled by pressing a membrane of polymer electrolyte between two supported CNT fiber sheets. The device is then encapsulated using a conventional plastic laminate machine to preserve a small pressure between the different layers. From a manufacturing point of view this process is attractive as it is compatible with roll-to-roll techniques, including those used in laminate composite fabrication. The process could be used to produce devices as large as 100cm$^2$ with electrodes fabricated in-house, and was also demonstrated with commercial CNT fiber sheets. Key for the reproducible fabrication of large-area devices is the high control over the synthesis and deposition of CNT fibres using the direct spinning floating catalyst CVD process, leading to electrodes with controlled thickness and uniform properties (composition, porosity, conductivity, etc).

Electrochemical measurements shows that the properties of devices with PE are very similar to those obtained using IL, and which can be taken as a benchmark to evaluate the efficiency of the EDLC assembly process. The maximum gravimetric capacitance, energy density and power densities realized in a PE-based device were 28 F g$^{-1}$, 11.4 Wh kg$^{-1}$ and 46 kW kg$^{-1}$, respectively. Power density was found to decrease with the reciprocal of electrode thickness. The linear dependence of galvanostatic discharge time with mass loading is in line with theoretical predictions for a symmetric EDLC with highly conductive electrodes and no micropore diffusion or micropore resistance. Commercial CNT fiber sheets also follow this trend. This emphasizes the need to specify electrode thickness when reporting power density values, or using the product of power density X mass loading = power/projected area, as a more meaningful figure of merit towards maximizing specific power density in real devices.

The EDLC devices could be repeatedly bent and folded 180° without degradation of their properties. In fact, bending produced reversible increases in energy density of as much as 25%. These results demonstrate the robustness of the polymer electrolyte/CNT fiber interface produced by our simple pressing method.

As a first step towards the use of these materials as structural EDLC, we have carried out electrochemical and mechanical testing of all-solid devices based on electrodes comprising highly oriented CNTs. As a pointer for the future, if the tensile properties of an individual filament can be realized in an electrode and the membrane thickness reduced by a forth, the resulting composite would have specific modulus and strength of approximately 0.37 GPa/SG and 18 GPa/SG, respectively. Once the current collector is removed and plastic laminate is replaced by a thinner equivalent, which so far looks entirely plausible, such composite would be able to store 1.0 Wh and deliver 2.0 kW per kg of *device*.

## 4. Experimental Section

**Synthesis of CNT fibers**

Carbon nanotubes fibers were synthesized by the direct spinning process from the gas-phase during growth of CNTs by floating catalyst chemical vapor deposition,[20] using ferrocene as iron catalyst, thiophene as a sulfur catalyst promoter and butanol or toluene as carbon sources. The reaction was carried out in hydrogen atmosphere at 1250 °C, using precursor feed rate of 5 mL h$^{-1}$ and a winding rate of 5 m min$^{-1}$ for butanol fibers and 2 mL h$^{-1}$ and 30 m min$^{-1}$ for toluene fibers. The CNT fibers were collected directly on aluminum substrate used as current collector for electrochemical characterization of the material. Mass loading of electrodes was varied from 0.16 to 2.40 mg cm$^{-2}$ by adjusting the time of CNT fibre collection. The film of fibers was consolidated by densification of as-spun samples with acetone, followed by room temperature drying.

The commercial CNT fiber samples were also produced by the direct spinning method. Samples 1 and 2 were supplied by Tortech Nano Fibers and INFRA, respectively. Sample 3 was produced by an undisclosed US-based supplier.

**Characterization of CNT fibers.**

Scanning (SEM) and transmission (TEM) electron microscopy images were obtained using a FIB-FEGSEM Helios NanoLab 600i (FEI) at 5 kV, a JEOL JEM 3000F TEM at 300kV and Tecnai F30 (FEI company) at 300 kV.

Gas adsorption measurements were carried out with $N_2$ at 77 K after drying the samples first at 100 ºC in an air atmosphere for 24 hours and later degassing at 300 ºC under vacuum for 3 hours using a Gemini VII 2390 Surface Area Analyzer (Micromeritics). The adsorption/desorption isotherms were obtained at 77K by allowing 6 sec for equilibration between each successive point. The pore size distribution was determined by the Barrett-Joyner-Halenda (BJH) analysis method applied to the nitrogen desorption isotherm data.

Thermogravimetric analysis (TGA) was performed using a TA Instruments TGAQ500 with a ramp of 10 °C min$^{-1}$ under nitrogen from 25 to 900 °C.

**Preparation of ionic liquid based polymer electrolyte membrane**

Ionic liquid based polymer electrolyte (PE) membranes were produced by first dissolving 3 g of PVDF-co-HFP (average M$_w$ ~400,000; Sigma Aldrich) and PYR$_{14}$TFSI (99.5% purity; Solvanoic) in 20 ml of acetone and stirring the mixture. Different mass ratio of polymer and IL in PE membrane (30:70, 40:60 and 50:50, 60:40) were obtained by adjusting the volume of IL in the initial solution. Transparent PE membranes with average thickness of 40 μm were produced by casting the solution, doctor blading. Solvent was evaporated at room temperature overnight, followed by annealing at 80 ºC under vacuum during 3 hours before fabrication of all-solid EDLCs.

**Assembly of self-standing all-solid EDLC devices.**

Before assembling EDLC devices all electrodes of CNT fibers were dried at 120ºC under vacuum during 3 hours. Fabrication of self-standing SCs consisted of sandwiching a pre-formed PE membrane between two CNT fiber electrodes of similar weight, introducing the composite into a plastic pouch film, applying a small pressure of 4 tons during 10 minutes with an uniaxial press (CARVER model 3853-0) and laminating the device using a conventional pouch laminator. This process produced excellent sealing. Samples were characterized at room temperature at least one hour after lamination in order to avoid possible spurious effect arising from the lamination process.

**Characterization of Mechanical properties**

The tensile tests were carried out with INSTRON 5966 dual column tabletop universal testing system using a gauge length of 15 mm and a test speed of 0.5 mm/min. Sample dimensions were determined with a digital micrometer. Specific properties obtained by dividing load by sample linear density gave similar values to those based on the cross-section stress normalized by specific gravity, thus confirming that the determination of sample cross-section is accurate. Fiber mass fraction was determined by weighing CNT fiber sheets and the PE membrane before assembly for each device. Volume fraction was calculated taking densities of 1.4 g cm$^{-3}$ and 1.78 g cm$^{-3}$ for PYR$_{14}$ TFSI and PVDF-co-HFP, respectively. For CNT fibers, the density in the composite was estimated as 0.7 g cm$^{-3}$ based on previous results infiltrating high molecular weight thermoplastics.[58]

**Characterization of Electrochemical Properties**

Individual properties of CNT fiber electrodes were characterized by cyclic voltammetry (CV) using a Biologic VMP multichannel potentiostatic–galvanostatic system and 3 electrode cell configuration with silver wire as the pseudo-reference and Pt mesh as the counter electrode. 1-butyl-1-methylpyrrolidinium bis(trifluoromethanesulfonyl) imide (PYR$_{14}$TFSI) ionic liquid was used as the electrolyte. Scan rates applied ranged from 5 to 1500 mV s$^{-1}$ and the voltage window used was up to 3.5 V. Specific capacitance ($C_s$) was obtained by integrating the area under CV curves and normalizing by mass of active material.

Symmetric liquid- and solid EDLCs were built by assembling two CNT fiber electrodes of similar weight and a commercial Celgard separator with pure IL or a PE membrane, in a two electrode Swagelok® cell. Self-standing all-solid devices were designed by sandwiching CNT fiber electrodes and PE membrane under constant pressure and then laminated by using plastic film and an ordinary laminating machine. Electrochemical behaviour of EDLCs was investigated by galvanostatic charge-discharge measurements using currents ranging from 1 to 20 mA cm$^{-2}$. Cycle stability tests of EDLCs consisted of over 10000 CD cycles at 5 mA cm$^{-2}$ and 3.5V in climatic chamber at fixed temperature of 25 °C. Capacitance of the full device ($C_{cell}$) was obtained from galvanostatic charge-discharge from the slope of the discharge curve as $C_{cell}$ = I/slope. In symmetric devices the specific capacitance of a single electrode can be extracted by using $C_s$ = 4 $C_{cell}$. Real energy density ($E_{real}$) and real power density ($P_{real}$) were calculated by integrating discharge curves according to the following equations:

$$E_{real} = I \int V dt \qquad (1)$$

$$P_{real} = \frac{E_{real}}{t_{dis}} \qquad (2)$$

**Supporting Information**
Supporting Information is available from the Wiley Online Library or from the author.

**Acknowledgements**
The authors acknowledge Tortech and Infra for kindly supplying CNT fiber sheet samples. Authors are grateful for assistance with gas adsorption measurements from L. Muñoz-Fernández and M. E. Rabanal, and for financial support provided by the European Union Seventh Framework Program under grant agreement 678565 (ERC-STEM) and FP7-People-Marie Curie Action-CIG (2012-322129 MUFIN), and from MINECO (MT2012-37552-C03-02, MAT2015-62584-ERC, MAT2015-64167-C2-1-R, RyC-2014-15115, Spain), and the Comunidad de Madrid MAD2D-CM Program (S2013/MIT-3007)

Received: ((will be filled in by the editorial staff))
Revised: ((will be filled in by the editorial staff))
Published online: ((will be filled in by the editorial staff))

**Table of content**

**All-solid large-are flexible electric double layer capacitors (EDLCs)** are assembled by a simple lamination method using CNT fiber sheets and ionic liquid-based polymer electrolyte (PE) membrane. Electric double layer capacitor devices of 100 cm$^2$ exhibit high electrochemical performance and excellent stability bent and folded. Highly-oriented CNT fiber/PE composite combine high power and high energy density with semi-structural mechanical properties.

**Keyword:** Carbon nanotube fibers, all-solid supercapacitor, energy storage, polymer electrolyte, structural composite

*Author(s), and Corresponding Author(s)\** Evgeny Senokos, Víctor Reguero, Laura Cabana, Jesus Palma, Rebeca Marcilla*, Juan Jose Vilatela*

**Title** Large-area, all-solid and flexible supercapacitors based on CNT fiber electrodes and polymer electrolytes.

**ToC figure**

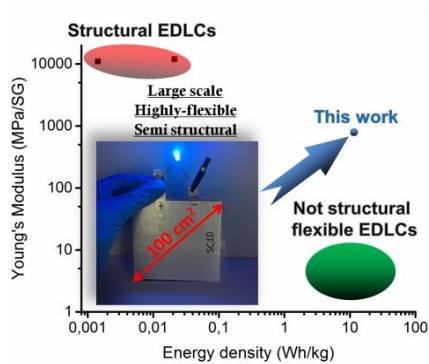

Supporting Information

**Title** Large-area, all-solid and flexible supercapacitors based on CNT fiber electrodes and polymer electrolytes.

*Author(s), and Corresponding Author(s)\** Evgeny Senokos, Víctor Reguero, Laura Cabana, Jesus Palma, Rebeca Marcilla*, Juan Jose Vilatela*

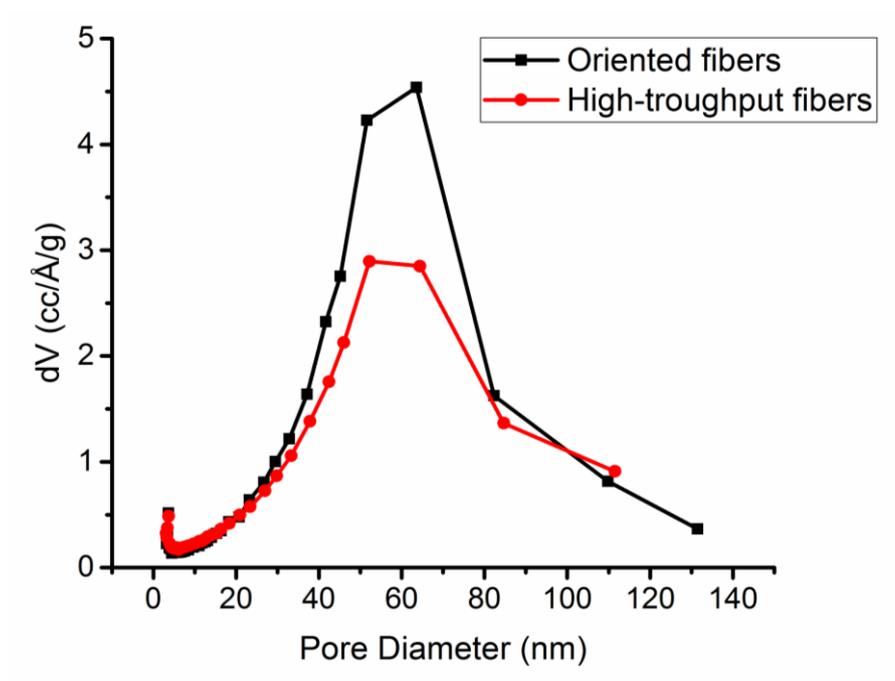

**Figure 1S.** Pore size distribution of high-throughput and oriented CNT fibers produced at different feed (5 mL h$^{-1}$ and 2 mL h$^{-1}$) and winding (5 m min$^{-1}$ and 40 m min$^{-1}$) rates, respectively.

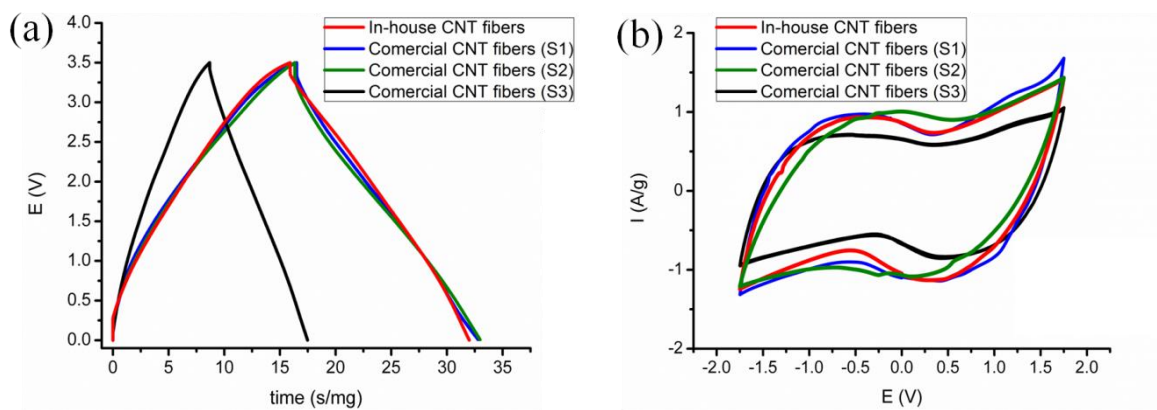

**Figure 2S.** a) Charge-discharge profiles of EDLC containing CNT fibers and b) Cyclic voltammograms in 3-electrode electrochemical cells of CNT fibers produced in-house (IMDEA) and at scale-up facilities (commercial suppliers S1, S2 and S3). All the experiments were performed in pure $PYR_{14}TFSI$.

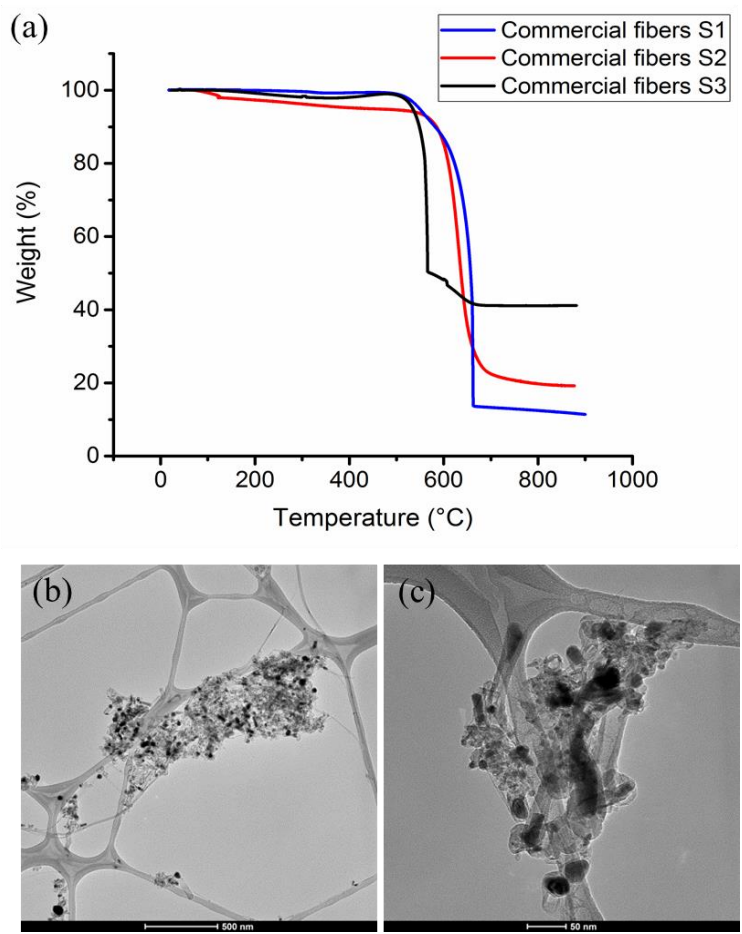

**Figure 3S.** a) TG curves comparing thermal behavior of commercial CNT fibers and b) TEM images of S3 sample showing high content of metal particles.

Sample S3 has a markedly lower gravimetric capacitance than the others studied in this work. This is attributed to a higher fraction of impurities, including an unusually high amount of residual catalyst of 41 wt. % determined by TGA (Figure 3Sa) and graphitic particulates observed by TEM (Figure 3Sb-c), both of which reduce the effective surface area for ion adsorption and thus the sample's gravimetric capacitance.

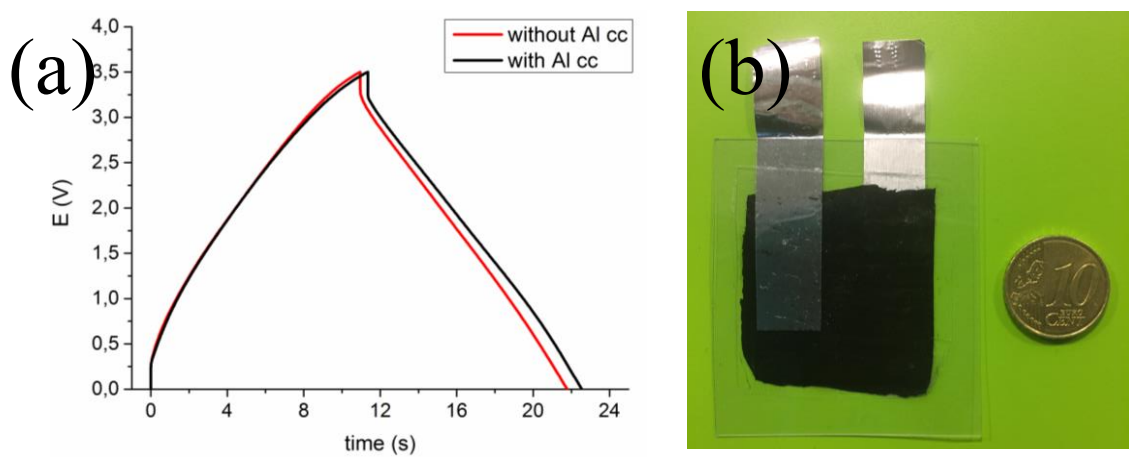

**Figure 4S.** All-solid EDLC without current collector. a) CD profiles comparing electrochemical performance of free-standing EDLCs fabricated with and without aluminium current collector and b) Photograph of all-solid EDLC device assembled without using aluminium foil support.

All-solid EDLC devices can be fabricated without using an Al foil as support/current collector (Figure 4S). The resulting CD profile shows a small ESR and indicates that the CNT fibre active materials has sufficiently high electrical conductivity to act also as current collector, thus producing a substantial reduction in device weight.

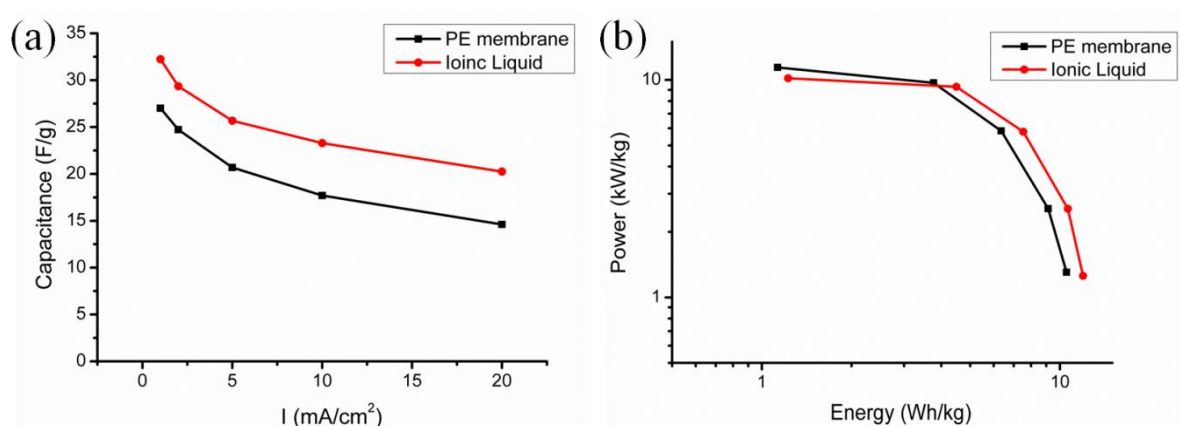

**Figure 5S.** a), b) Capacitance and Ragone plots comparing all-solid EDLC based on polymer electrolyte membrane and EDLC based on pure IL.

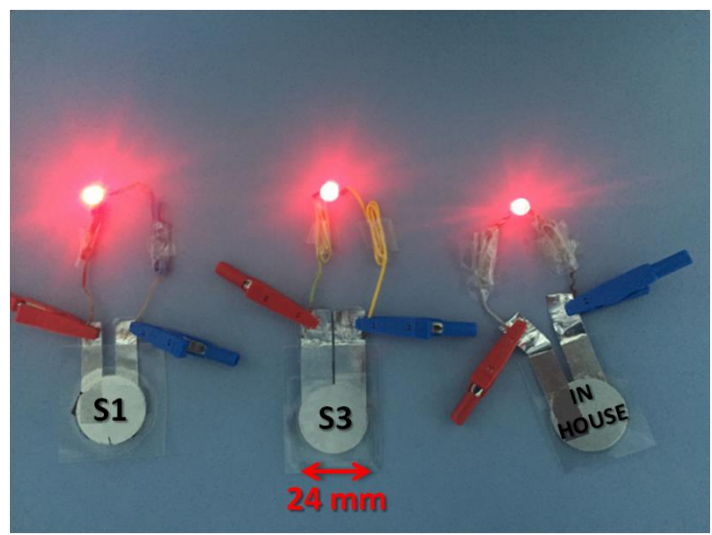

**Figure 6S.** All-solid EDLC devices of 4.5 cm$^2$ power a red LED, including devices based on electrodes produced in house and on electrodes from commercial CNT fiber suppliers.

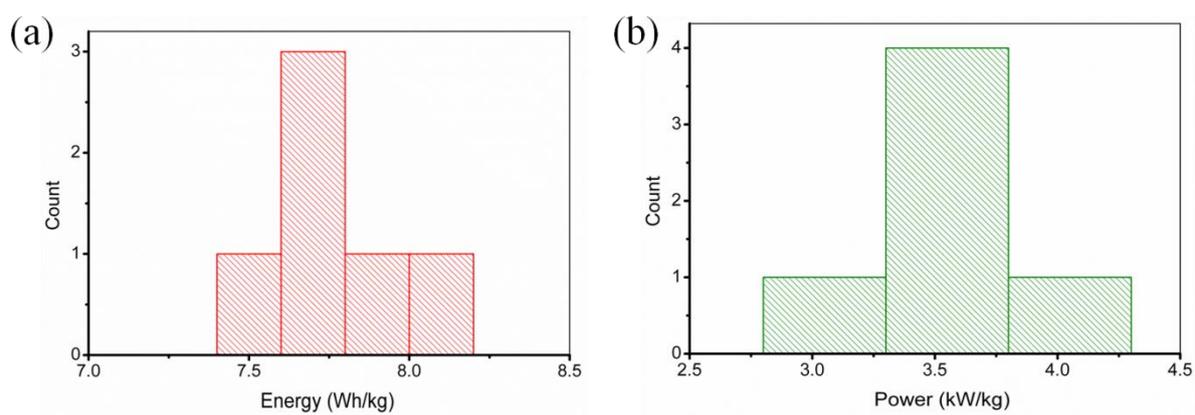

**Figure 7S.** Histograms showing the reproducibility of a) Energy density and b) Power density of 100 cm$^2$ free-standing all-solid EDLCs.

32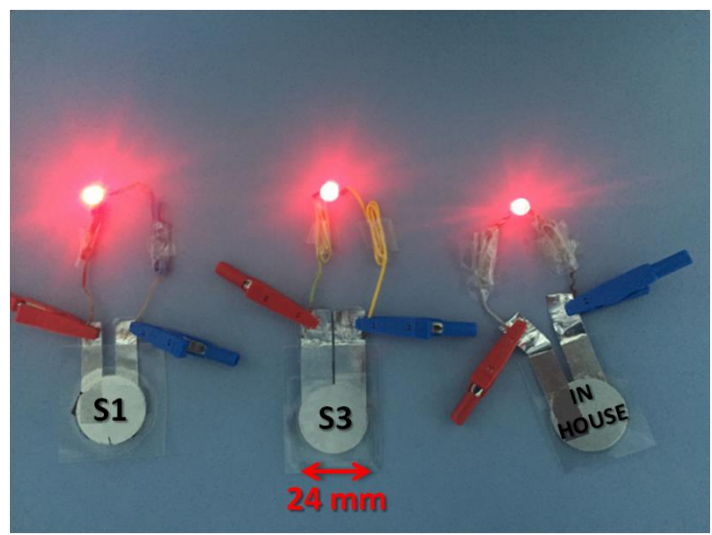

**Figure 6S.** All-solid EDLC devices of 4.5 cm$^2$ power a red LED, including devices based on electrodes produced in house and on electrodes from commercial CNT fiber suppliers.

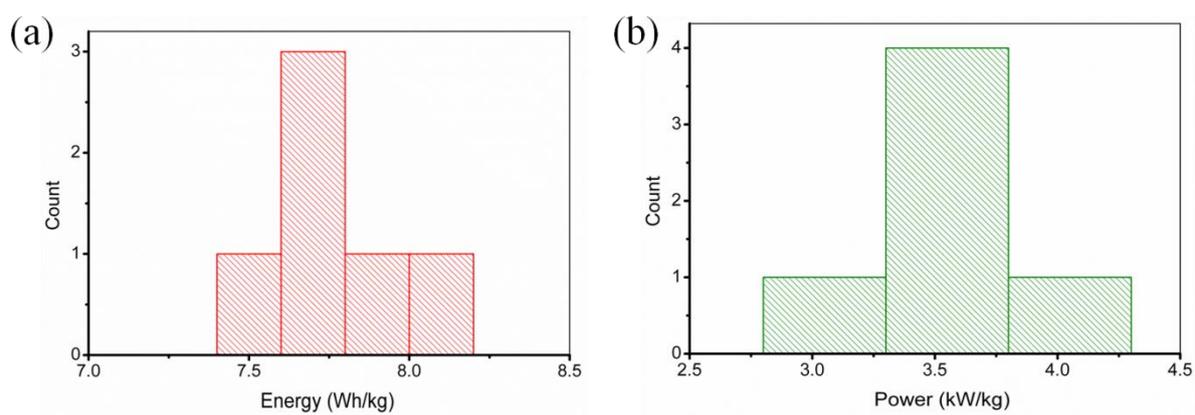

**Figure 7S.** Histograms showing the reproducibility of a) Energy density and b) Power density of 100 cm$^2$ free-standing all-solid EDLCs.

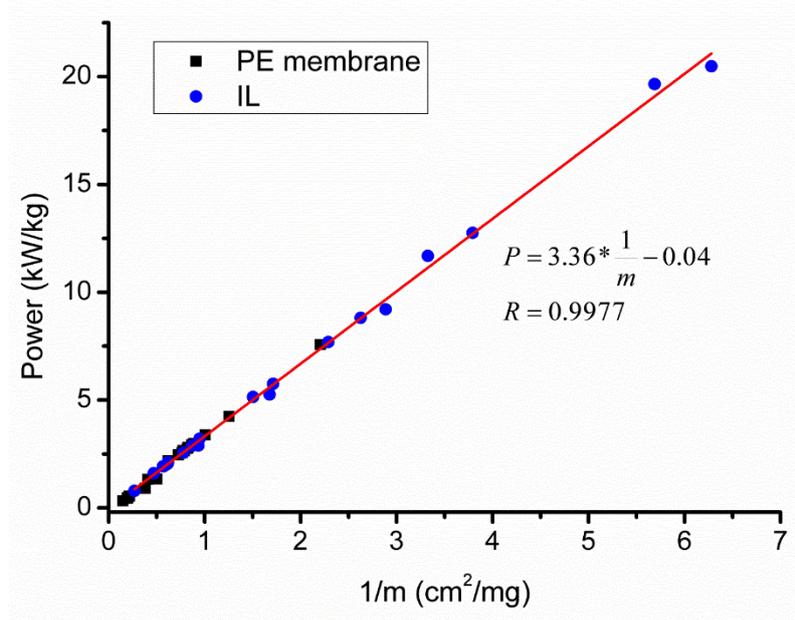

**Figure 8S.** Power density of EDLCs with CNT fiber sheet electrodes obtained from CD measured at 2 mAcm$^{-2}$ plotted versus reciprocal electrode mass loading showing identical linear dependence both for liquid state (IL) and all-solid (PE membrane) EDLCs.

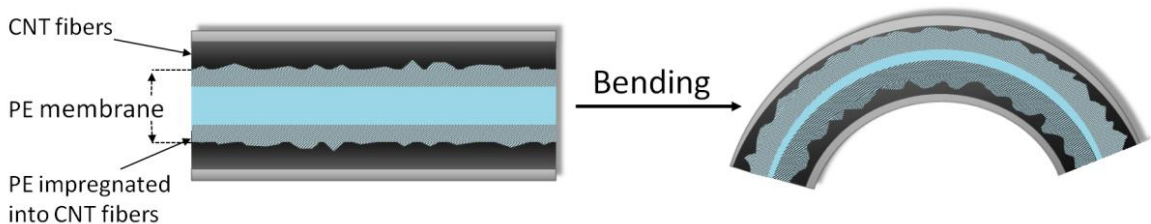

**Figure 9S.** Scheme of laminated EDLC's cross-section demonstrating changes in a degree of PE membrane impregnation before and after bending the device.

Under bending, the laminar structure will experience a compressive stress normal to the plane. Because of the viscoelastic nature of the polyelectrolyte, this stress can induce additional infiltration into the porous CNT fibre structure. The formation of new polyelectrolyte/CNT fibre interface implies that a larger charge can be stored and thus that energy density increases. However, power density remain unchanged, since the increase in energy comes with a proportional increase in charge/discharge time, as discussed in the text for electrodes with different thickness (see Figure 2).